\documentclass[twocolumn]{aastex631}

\newcommand\cd{d$^{-1}$}

\received{2021 July 30}
\accepted{2021 August 10}
\published{}

\submitjournal{ApJL}

\shorttitle{Discovery of pulsations in PG 1144+005}
\shortauthors{Sowicka et al.}
\graphicspath{{./}{figures/}}

\begin{document}

\title{The Missing Link? Discovery of Pulsations in the Nitrogen-rich PG 1159 Star PG 1144+005}

\correspondingauthor{Paulina Sowicka, Gerald Handler}
\email{paula@camk.edu.pl, gerald@camk.edu.pl}

\author[0000-0002-6605-0268]{Paulina Sowicka}
\affiliation{Nicolaus Copernicus Astronomical Center, Polish Academy of Sciences, ul. Bartycka 18, PL-00-716, Warszawa, Poland}

\author[0000-0001-7756-1568]{Gerald Handler}
\affiliation{Nicolaus Copernicus Astronomical Center, Polish Academy of Sciences, ul. Bartycka 18, PL-00-716, Warszawa, Poland}

\author[0000-0003-3947-5946]{David Jones}
\affiliation{Instituto de Astrof\'isica de Canarias, E-38205 La Laguna, Tenerife, Spain}
\affiliation{Departamento de Astrof\'isica, Universidad de La Laguna, E-38206 La Laguna, Tenerife, Spain}

\author{Francois van Wyk}
\affiliation{South African Astronomical Observatory, P.O. Box 9, Observatory, 7935 Cape, South Africa}



\begin{abstract}

Up to 98\% of all single stars will eventually become white dwarfs - stars that link the history and future evolution of the Galaxy, and whose previous evolution is engraved in their interiors. Those interiors can be studied using asteroseismology, utilizing stellar pulsations as seismic waves. The pulsational instability strips of DA and DB white dwarf stars are pure, allowing the important generalization that their interior structure represents that of all DA and DB white dwarfs. This is not the case for the hottest pulsating white dwarfs, the GW Vir stars: only about 50\% of white dwarfs in this domain pulsate. Several explanations for the impurity of the GW Vir instability strip have been proposed, based on different elemental abundances, metallicity, and helium content. Surprisingly, there is a dichotomy that only stars rich in nitrogen, which by itself cannot cause pulsation driving, pulsate -- the only previous exception being the nitrogen-rich non-pulsator PG~1144+005.
Here, we report the discovery of pulsations in PG~1144+005 based on new observations. 
We identified four frequency regions: 40~\cd, 55~\cd, 97~\cd, and 112~\cd with low and variable amplitudes of about $3-6$~mmag and therefore confirm the nitrogen dichotomy. As nitrogen is a trace element revealing the previous occurrence of a very late thermal pulse (VLTP) in hot white dwarf stars, we speculate that it is this VLTP that provides the interior structure required to make a GW Vir pulsator.

\end{abstract}



\section{Introduction} \label{sec:intro}
White dwarf stars are the most common endpoint of stellar evolution with up to 98\% of all single stars eventually reaching this phase. Despite their importance, detailed knowledge of the interior structure of only a limited number of white dwarfs is available. White dwarf stars pulsate in certain zones of instability along their cooling curves that define the three classical types of white dwarf pulsators (e.g. \citealt{2008ARA&A..46..157W}) within the growing family of (pre-)white dwarf pulsators \citep{2019A&ARv..27....7C}. The first pulsational instability strip entered by a post-AGB star is that of the GW~Vir stars, followed by the domains of the DB and DA white dwarf pulsators. White dwarfs residing in the GW~Vir domain span a large range of effective temperatures ($\approx 75 - 250$ kK) and surface gravities (log g $\approx 5.5 - 8$). Some stars show photometrically detectable pulsations with periods as short as a few minutes due to nonradial gravity modes driven by the $\kappa$ mechanism associated with the partial ionization of the K-shell electrons of carbon and/or oxygen in the envelope \citep{1983ApJ...268L..27S, 1984ApJ...281..800S}. 

GW~Vir pulsators are distinguished by their spectra: most are of the PG~1159-type but there are also [WC]-types (central stars of planetary nebulae with Wolf-Rayet spectra of the carbon sequence, e.g., \citealt{1998MNRAS.296..367C}). Such stars are thought to be formed as a result of a ``born-again'' episode (a very late thermal pulse [VLTP] experienced by a hot white dwarf during its early cooling phase) or a late thermal pulse (LTP) that occurs during the post-AGB evolution when H burning is still active (e.g., \citealt{1983ApJ...264..605I}, \citealt{2001Ap&SS.275....1B}, \citealt{2006A&A...454..845M}). They are supposed to be the main progenitors of H-deficient white dwarfs, which makes them important to study in the context of stellar evolution. PG~1159 stars exhibit He-, C- and O-rich surface abundances, but strong variations of the He/C/O ratio were found from star to star (\citealt{1998A&A...334..618D}; \citealt{2001Ap&SS.275...27W}), as were traces of other, heavier elements. The presence of H (such stars are classified as ``hybrid-PG~1159'') together with a N dichotomy (N-rich, about 1\% in mass and N-poor, below about 0.01\% in mass, stars) are tracers of the evolutionary history, i.e. when the progenitor experienced the final thermal pulse \citep{2001Ap&SS.275...15H,2008ASPC..391..109W}. 
The variety of surface abundance patterns observed in PG~1159 stars poses indeed a challenge to the theory of stellar evolution, but its understanding may be key in revealing their evolutionary history.

The instability strips of the pulsating DA and DB white dwarfs are believed to be pure, i.e. all stars within their respective borders do pulsate. Consequently, the pulsators are otherwise normal white dwarfs and their interiors – that can be studied using the technique of asteroseismology – represent the interiors of all white dwarfs. This is not the case for the PG~1159 stars, as the GW~Vir instability strip is not pure; according to the literature (e.g., \citealt{2004ApJ...610..436Q}) only about 50\% of them pulsate. This raises the questions of what separates the pulsators from the non-pulsators and whether there are fundamental differences in their interior structures and thus evolutionary histories. 

The observed nitrogen dichotomy, i.e. N-rich stars are pulsators, whereas N-poor stars are all non-pulsators \citep{1998A&A...334..618D}, suggests that N may play a role in pulsation driving, despite its rather small abundance even in N-rich stars. \citet{2007ApJS..171..219Q} refuted this idea finding that predominantly a high O abundance is responsible for pulsation driving.
The whole picture, however, is more complicated, involving the physical parameters and chemical compositions of PG~1159 stars, including metallicity and the role of helium ``poisoning'' the driving of pulsations. Interesting in this context then is the part played by nitrogen as a tracer of the previous evolutionary history of GW Vir stars.  
Is a VLTP a necessity for achieving the chemical mixture required to destabilize a star to develop pulsations?
This would allow the important conclusion that the GW Vir stars have a fundamentally different evolutionary history than the non-pulsators.
In this regard, one final culprit however still remained: PG 1144+005. Ever since the detection of strong \ion{N}{5} emission lines in its spectrum \citep{1991A&A...247..476W} and the realization that all pulsating PG~1159 stars are nitrogen rich \citep{1998A&A...334..618D} it was the only known N-rich PG~1159 star that was never discovered to pulsate.

PG~1144+005 ($T_{\mathrm{eff}}=150\,000\pm15\,000$~K, $\log g=6.5\pm0.5$, \citealt{1991A&A...247..476W}) is a G=15.1734~mag star \citep{2018A&A...616A...1G} found in the Palomar-Green (PG) Survey by the UV excess \citep{1986ApJS...61..305G}. A number of authors observed PG~1144+005 since then, over the last 30~years. In the pioneering search for extremely hot pulsating stars, \citet{1987fbs..conf..231G,1987ApJ...323..271G} did not find any variability consistent with pulsations among the program stars, including PG~1144+005, regrettably not providing the detection limit. 
Over a decade later, \citet{2000BaltA...9..395S} observed it twice, 
with a null result. Unfortunately, again, also these authors did not note the detection limit of these observations. Finally, \citet{2003ASPC..292..237S} revisited the object and observed it for four nights on the 1-m telescope at Piszkestet\"o. Data collected over 5.6~hours showed a light curve with no visible variability and a semi-amplitude of about 0.02~mag. The lack of variability was then reflected in the Fourier amplitude spectrum over the range from about $50$~\cd to about $700$~\cd, where no peak exceeded an amplitude of 4~mmag. No further time-series observations of the star have been reported since.


\section{Discovery observations with GTC}
We included PG~1144+005 in a sample of PG~1159 stars selected for a survey for variability (Sowicka P. et al. 2021, in preparation). The observations were carried out on 2018 January 17 with the 10-m Gran Telescopio Canarias (GTC) equipped with OSIRIS in one observing block as a filler target.
OSIRIS consists of a mosaic of two CCDs 2048 x 4096 pixels each and has an unvignetted field of view (FOV) of 7.8 x 7.8~arcmin. 
We used a Sloan~r' filter with an exposure time of 6~seconds and standard readout time, resulting in a duty cycle of about 29~seconds, over the 1.1~hours length of observations. 
The data were reduced using standard Astropy \citep{2013A&A...558A..33A,2018AJ....156..123A} \textsc{ccdproc} \citep{matt_craig_2017_1069648} routines consisting of bias subtraction, flat-field and gain correction.  

PG~1144+005 is a challenging object for differential photometry. Because of the lack of suitable nearby comparison stars in the typical small field of view of relatively large telescopes, we used a ``master'' comparison star created as a sum of flux from three stars after making sure they are photometrically constant: MGC~24209, MGC~24219, MGC~24226, to improve the S/N in the light curve as all those stars are at least 1~mag fainter than the target in $r'$.

\begin{figure*}
\includegraphics[width=\textwidth]{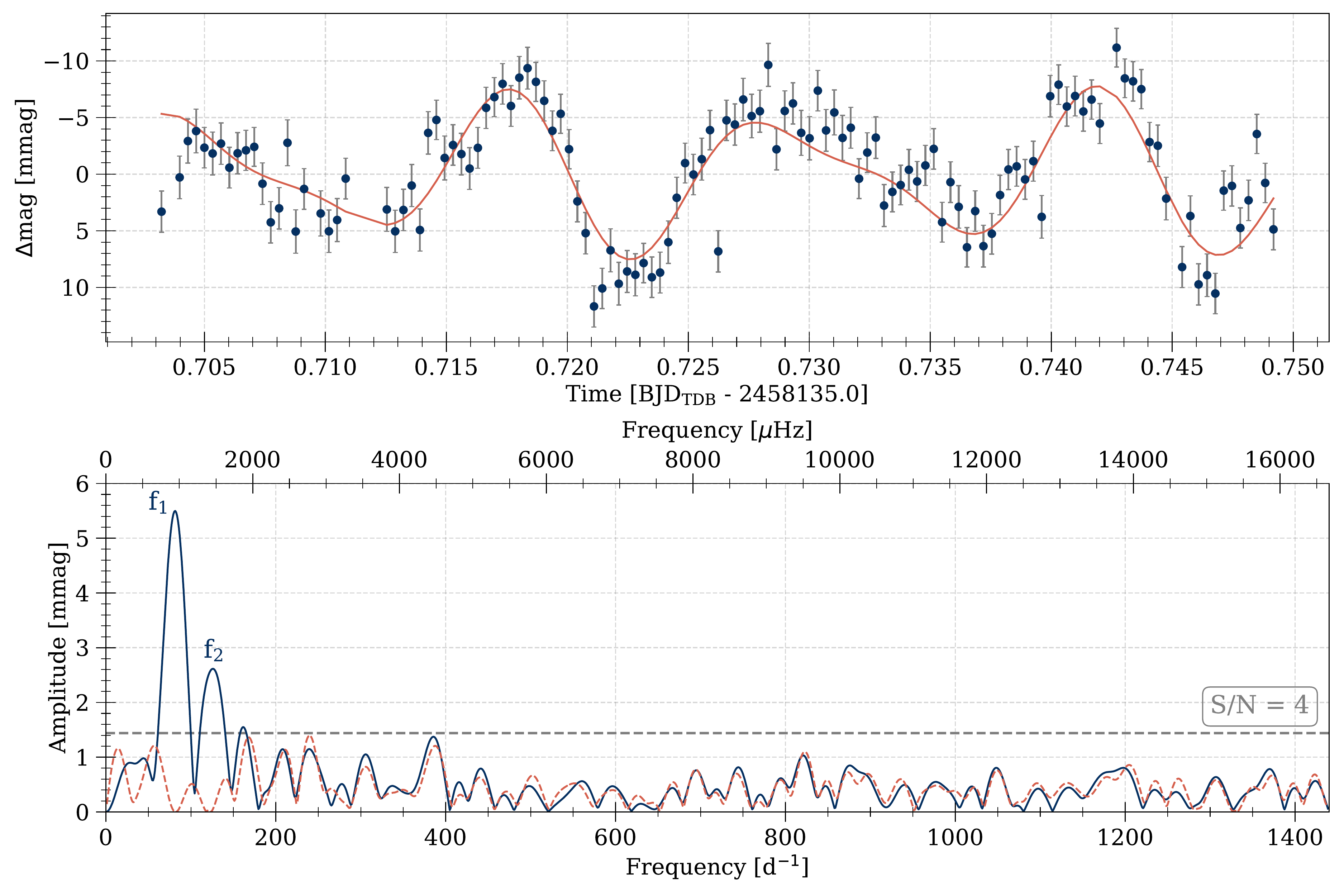}
 \caption{Top: GTC differential light curve of PG~1144+005 (blue circles) with a temporary model fit calculated using $f_1$ and $f_2$ shown for clarity. Bottom: Fourier amplitude spectrum (blue solid line) with significance criterion of S/N = 4 (gray dashed line). The two extracted frequencies are labeled, while the residual amplitude spectrum after prewhitening of these two modes is shown as orange dashed line.
\label{fig:gtclcft}}
\end{figure*}

We performed aperture photometry using our own procedures utilizing scaled adaptive aperture sizes to the seeing conditions, as described in \citet{2018MNRAS.479.2476S}, in this case characterized by a scaling factor of $1.5\times$FWHM for each frame. The differential light curve was then corrected for differential color extinction. Fig.~\ref{fig:gtclcft} shows the final light curve with clear variability, and its Fourier amplitude spectrum up to the Nyquist frequency of $1439$~\cd, calculated using \textsc{Period04}~\citep{2005CoAst.146...53L}. 

The semi-amplitude is about 5~mmag in the light curve, while the median noise level of the Fourier amplitude spectrum is 0.36~mmag (calculated in the range $400-1439$~\cd). We identified two peaks, f$_1=81.60\pm0.82$~\cd with amplitude A$_{1}=5.37\pm0.37$~mmag (S/N ratio of $14.92$) and f$_2=124.036\pm1.70$~\cd with A$_2=2.60\pm0.37$~mmag (S/N ratio of $7.22$). 

This is good evidence that the only remaining N-rich PG~1159 star, PG~1144+005, is a pulsating star and therefore belongs to the GW Vir family. However, the short light curve and the relatively long variability periods called for confirmation.


\section{Follow-up SAAO observations}

Follow-up observations were carried out on five nights in 2021 May 6, 8, 9, 10, 11. We observed PG~1144+005 using the 1-m Lesedi telescope located at South African Astronomical Observatory (SAAO), equipped with Sutherland High Speed Optical Camera (SHOC; \citealt{2013PASP..125..976C}) instrument SHA. 
Thanks to Lesedi's 5.7 x 5.7~arcmin FOV, the $G=12.8$ magnitude star MGC~24274 (saturated in GTC observations) could be used as the primary comparison star. PG~1144+005 was visible for about 5~hours in the first half of the nights and we collected 24.06~hours of data. We varied the exposure time during the nights to adjust to the changing atmospheric conditions and airmass, to avoid saturation of the comparison star, and to mitigate possible Nyquist frequency ambiguities. Table~\ref{tab:SAAOlog} shows the detailed log of observations. 

\begin{table}[]
\centering
\begin{tabular}{llcc}
\hline
Night & UTC start & \multicolumn{1}{l}{Exp. time [s]} & \multicolumn{1}{l}{$\#$ of frames} \\ \hline
2021-05-06 & 18:41:40.360181 & 20      & 120  \\
           & 19:38:46.344030 & 10      & 1206 \\ \hline
2021-05-08 & 17:05:31.746625 & 10      & 327  \\
           & 18:02:09.824185 & 20      & 115  \\
           & 19:29:37.096074 & 10      & 364  \\
           & 20:30:33.424220 & 9       & 68   \\
           & 20:41:05.839185 & 8       & 790  \\
           & 22:26:44.605522 & 10      & 152  \\ \hline
2021-05-09 & 17:32:45.588846 & 10      & 170  \\
           & 18:01:49.315274 & 8       & 1090 \\
           & 20:27:54.069439 & 7       & 540  \\
           & 21:31:17.963203 & 8       & 190  \\
           & 21:56:52.790226 & 10      & 65   \\
           & 22:07:57.782761 & 9       & 268  \\ \hline
2021-05-10 & 16:53:12.174974 & 8       & 300  \\
           & 17:40:50.857600 & 8       & 300  \\
           & 18:35:59.625951 & 8       & 844  \\
           & 20:28:58.190206 & 7       & 600  \\
           & 21:39:17.597373 & 8       & 486  \\ \hline
2021-05-11 & 16:57:30.036797 & 10      & 600  \\
           & 18:37:45.304383 & 8       & 1200 \\
           & 21:19:29.860689 & 10      & 330  \\
           & 22:14:42.561065 & 12      & 60   \\ \hline
\end{tabular}
\caption{Journal of time-series photometric observations of PG~1144+005 in May 2021. Some parts of data were removed because of bad quality.}
\label{tab:SAAOlog}
\end{table}

\begin{figure}
\centering
\includegraphics[height=0.95\textheight]{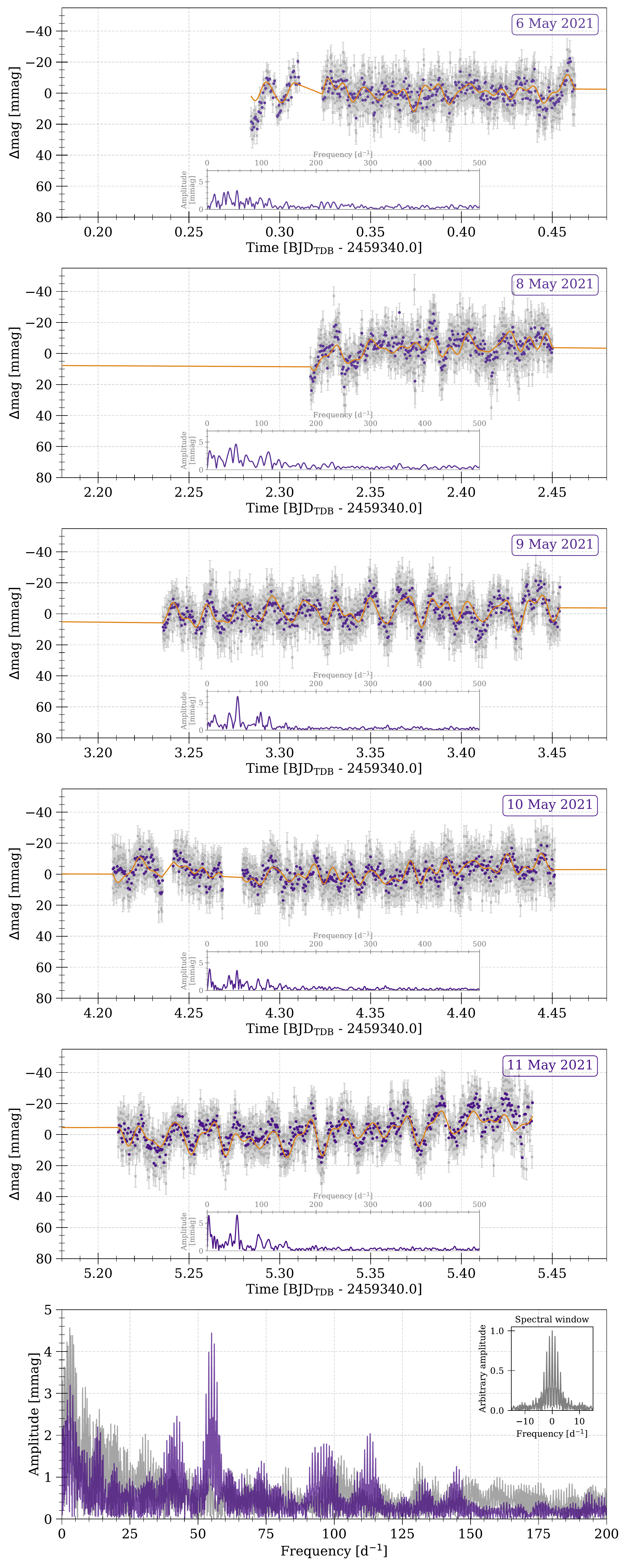}
 \caption{Top 5 panels: nightly light curves with amplitude spectra as insets. Gray `x' with error bars - original data points, filled purple circles - 40-s binned data points. A fit from \textsc{Period04} is shown as orange solid line to indicate variability. Bottom panel: Combined amplitude spectra of the run as semi-transparent purple and grey solid lines for PG~1144+005 and the comparison star, respectively. The spectral window is shown in inset.
\label{fig:saao_lcft}}
\end{figure}

We performed data reduction (including bias, flat-field, and gain correction) and differential aperture photometry using \textsc{TEA-Phot} \citep{2019A&A...629A..21B}. \textsc{TEA-Phot} is a data reduction and photometry package especially designed to work with data cubes from the SHOC instruments. It uses adaptive elliptical apertures to extract photometry, where the optimum major and minor axes of the apertures are calculated for each frame from an initial guess from the user based on examination of the displayed curve of growth for the target and comparison star separately.
We then corrected each combined nightly light curve for differential colour extinction and removed outlying points (3.5-$\sigma$ clipping) and bad-quality parts of data (observations through thick clouds). Fig.~\ref{fig:saao_lcft} shows the final light curves from each night, and binned in 40-s over-plotted to show the variability more clearly. 

The best quality data (in terms of observing conditions) come from the end of the run. The time base of the whole run is $\Delta$T=5.1554~d. The frequency resolution needed to resolve the modes and determine the amplitudes and phases correctly is calculated as $\Delta$f=1.5/$\Delta$T \citep{1978Ap&SS..56..285L} and equals to $\Delta$f=0.29096~\cd. We first calculated Fourier amplitude spectra of each nightly light curve separately. Then we combined the data and calculated the amplitude spectrum of the whole run, together with the spectral window (a single sine wave of arbitrary, constant amplitude - in our case 1 - sampled at the times of our time series data, which shows the aliasing pattern around each frequency in the data). The nightly light curves, amplitude spectra and the spectral window are shown in Fig.~\ref{fig:saao_lcft}. 
Frequencies, amplitudes and phases were determined by simultaneously fitting a nonlinear least-squares solution to the data using \textsc{Period04}~\citep{2005CoAst.146...53L} and the formal solution is presented in Table~\ref{tab:frequencies}, where we omitted frequencies below 10~\cd that we judged to have originated in the Earth's atmosphere.

\begin{table}[]
\centering
\begin{tabular}{lDDD}
\hline
& \multicolumn{2}{c}{Frequency} & \multicolumn{2}{c}{Frequency} & \multicolumn{2}{c}{Amplitude} \\ 
& \multicolumn{2}{c}{[\cd]} & \multicolumn{2}{c}{[$\mu$Hz]} & \multicolumn{2}{c}{[mmag]} \\ \hline
\decimals
$\nu_1$  &  54.963(3)  &     636.15(3) &  4.39(12) \\ 
$\nu_2$  &  56.528(6)  &     654.26(7) &  2.00(12) \\
$\nu_3$  &  113.163(7) &    1309.76(8) &  1.98(12) \\ 
$\nu_4$  &  42.219(7)  &     488.65(8) &  1.96(12) \\ 
$\nu_5$  &  39.542(7)  &     457.66(8) &  1.90(12) \\ 
$\nu_6$\footnote{The +1\cd ~alias of $\nu_6$ would correspond to $\nu_1+\nu_4$ within the errors.}  &  96.147(7)  &    1112.81(8) &  1.88(12) \\ 
$\nu_7$  &  91.945(8)  &    1064.18(9) &  1.61(12) \\ 

\hline
\end{tabular}
\caption{The formal frequency solution using the combined SAAO data set with analytical uncertainties. Frequencies below 10~\cd are omitted. }
\label{tab:frequencies}
\end{table}

In the new data we detect the presence of peaks at the same frequency ranges as in our exploratory GTC run, with the main pulsation modes grouped around 40~\cd, 55~\cd, 97~\cd, and 112~\cd. With the caveat that the two higher-frequency groups may be combination frequencies of the two lower-frequency ones, the detected frequencies are consistent with g-mode pulsations excited by the $\kappa$-mechanism due to partial ionization of carbon and/or oxygen in the envelopes of PG~1159 stars. The frequencies correspond to periods of about 700 to 3000~s, in agreement with the period ranges of other known variable PG~1159 stars (e.g. Fig.~7 in \citealt{2006A&A...458..259C}). We also find that the amplitude spectrum of PG~1144+005 is variable, with amplitudes of the modes changing between consecutive nights, but the frequencies remaining the same within the temporal resolution. It can be most clearly seen in the case of the region around 55~\cd in the insets in Fig.~\ref{fig:saao_lcft}, where the amplitudes change between less than 4~mmag to over 6~mmag. This kind of temporal variability has been shown for other GW Vir stars, either as a highly variable amplitudes and/or frequencies, or even as a complete disappearance of the pulsations for a period of time. A more conservative interpretation is beating of closely-spaced pulsation frequencies.

Also, according to \citet{2009ApJ...701.1008C}, PG~1144+005 lies in the overlapping region in the HR diagram where both $\kappa-$ and $\epsilon$-mechanism can operate. We therefore inspected our nightly amplitude spectra for the presence of high-frequency, $\epsilon$-driven modes. We did not find any peaks above the detection threshold of 1~mmag for frequencies above 150~\cd.

Even though we do not attempt for creating an asteroseismic model for PG~1144+005, our most important result is the confirmation of the variability preliminarily detected in our GTC run.

\section{Summary and discussion}
We obtained new observations of the N-rich PG~1159 star PG~1144+005 allowing for the discovery of long-sought pulsations in this star. The first, short run from the GTC allowed us to estimate the expected frequencies and amplitudes ($80-130$~\cd and $3-6$~mmag). Follow-up observations over 5 nights at SAAO clearly showed low-amplitude multi-periodic pulsations of PG~1144+005. The pulsations mainly appeared in four regions: 40~\cd, 55~\cd, 97~\cd, and 112~\cd, while the amplitudes were variable over the course of observations. The detected variability is consistent with g-mode pulsations excited in PG~1159 stars.

The g-mode pulsations of PG~1159 stars are low-amplitude high-frequency in nature. Most PG~1159 stars are faint, making required high-speed and high-quality follow-up observations challenging. To resolve individual modes and rotational splittings, the time base of a few days is needed (depending on the rotation period). The application of multisite campaigns for the study of these stars, as well as other pulsating white dwarfs and subdwarfs, has already been successfully shown, although only for a handful of brightest targets. Usually many large telescopes are needed, and the observing time hard to get. Are the future space telescopes for faint blue stars the only way to study the faintest ones (especially the PG~1159 stars discovered in SDSS)? 

Our discovery of the pulsations in PG~1144+005 provides the last missing piece to the current picture of the excitation mechanism and abundance patterns in PG~1159 stars. Is the N-problem thus solved, and the whole picture is complete, or will we find more N-rich non-pulsators or N-poor pulsators, if we push the detection limit down?

Out of 55 known PG~1159 stars only fourteen objects have the nitrogen abundance, or its upper limit, assessed. These are: the N-rich stars PG~1144+005, PG~1159-035, PG~2131+066, PG~1707+427, PG~0122+200, Abell~43, and N-poor stars PG~1520+525, PG~1424+535, HS~1517+7403, MCT~0130-1937, HS~0704+6153, H~1504+65, RXJ~0439.8-6809, NGC~7094. Prerequisites for this type of analysis were spectroscopic observations in (far-)ultraviolet and model atmospheres for such high effective temperatures. Therefore the stars with nitrogen abundance assessed are those having good-quality HST spectra available, with nitrogen lines ideally not blended with strong interstellar absorption. In only a few cases additional optical N lines could have been used. Even though the evolutionary link to the nitrogen abundance appears to hold, this is still a small number statistics considering the total number of known PG~1159 stars. Detailed assessment of element abundances for a larger sample of stars is needed.

For the time being, however, there is a clear separation: PG 1159 stars with significant amounts of nitrogen in the atmosphere pulsate, the others do not. This is evidence that pulsating and nonpulsating PG 1159 stars have different evolutionary histories. It seems necessary that a star has to undergo a VLTP to become a pulsator, and nitrogen is a tracer of this history.


\begin{acknowledgments}
We thank Bruno Steininger for discussions and for supplying his observational data. PS and GH acknowledge financial support by the Polish NCN grant 2015/18/A/ST9/00578. DJ acknowledges support from the Erasmus+ programe of the European Union under grant No. 2020-1-CZ01-KA203-078200. This paper uses observations made at the South African Astronomical Observatory (SAAO). Based on observations made with the Gran Telescopio Canarias (GTC), installed in the Spanish Observatorio del Roque de los Muchachos of the Instituto de Astrof\'isica de Canarias, in the island of La Palma. 
\end{acknowledgments}

%

\vspace{5mm}
\facilities{GTC (OSIRIS), SAAO: Lesedi (SHOC)}


\software{Astropy \citep{2013A&A...558A..33A,2018AJ....156..123A},
          \textsc{ccdproc} \citep{matt_craig_2017_1069648},
          Matplotlib \citep{Hunter:2007},
          \textsc{Period04} \citep{2005CoAst.146...53L},          
          \textsc{TEA-Phot} \citep{2019A&A...629A..21B}
          }







\bibliography{PG1144_v4}{}

\begin{thebibliography}{}
\expandafter\ifx\csname natexlab\endcsname\relax\def\natexlab#1{#1}\fi
\providecommand{\url}[1]{\href{#1}{#1}}
\providecommand{\dodoi}[1]{doi:~\href{http://doi.org/#1}{\nolinkurl{#1}}}
\providecommand{\doeprint}[1]{\href{http://ascl.net/#1}{\nolinkurl{http://ascl.net/#1}}}
\providecommand{\doarXiv}[1]{\href{https://arxiv.org/abs/#1}{\nolinkurl{https://arxiv.org/abs/#1}}}

\bibitem[{{Astropy Collaboration} {et~al.}(2013){Astropy Collaboration},
  {Robitaille}, {Tollerud}, {Greenfield}, {Droettboom}, {Bray}, {Aldcroft},
  {Davis}, {Ginsburg}, {Price-Whelan}, {Kerzendorf}, {Conley}, {Crighton},
  {Barbary}, {Muna}, {Ferguson}, {Grollier}, {Parikh}, {Nair}, {Unther},
  {Deil}, {Woillez}, {Conseil}, {Kramer}, {Turner}, {Singer}, {Fox}, {Weaver},
  {Zabalza}, {Edwards}, {Azalee Bostroem}, {Burke}, {Casey}, {Crawford},
  {Dencheva}, {Ely}, {Jenness}, {Labrie}, {Lim}, {Pierfederici}, {Pontzen},
  {Ptak}, {Refsdal}, {Servillat}, \& {Streicher}}]{2013A&A...558A..33A}
{Astropy Collaboration}, {Robitaille}, T.~P., {Tollerud}, E.~J., {et~al.} 2013,
  \aap, 558, A33, \dodoi{10.1051/0004-6361/201322068}

\bibitem[{{Astropy Collaboration} {et~al.}(2018){Astropy Collaboration},
  {Price-Whelan}, {Sip{\H{o}}cz}, {G{\"u}nther}, {Lim}, {Crawford}, {Conseil},
  {Shupe}, {Craig}, {Dencheva}, {Ginsburg}, {VanderPlas}, {Bradley},
  {P{\'e}rez-Su{\'a}rez}, {de Val-Borro}, {Aldcroft}, {Cruz}, {Robitaille},
  {Tollerud}, {Ardelean}, {Babej}, {Bach}, {Bachetti}, {Bakanov}, {Bamford},
  {Barentsen}, {Barmby}, {Baumbach}, {Berry}, {Biscani}, {Boquien}, {Bostroem},
  {Bouma}, {Brammer}, {Bray}, {Breytenbach}, {Buddelmeijer}, {Burke},
  {Calderone}, {Cano Rodr{\'\i}guez}, {Cara}, {Cardoso}, {Cheedella}, {Copin},
  {Corrales}, {Crichton}, {D'Avella}, {Deil}, {Depagne}, {Dietrich}, {Donath},
  {Droettboom}, {Earl}, {Erben}, {Fabbro}, {Ferreira}, {Finethy}, {Fox},
  {Garrison}, {Gibbons}, {Goldstein}, {Gommers}, {Greco}, {Greenfield},
  {Groener}, {Grollier}, {Hagen}, {Hirst}, {Homeier}, {Horton}, {Hosseinzadeh},
  {Hu}, {Hunkeler}, {Ivezi{\'c}}, {Jain}, {Jenness}, {Kanarek}, {Kendrew},
  {Kern}, {Kerzendorf}, {Khvalko}, {King}, {Kirkby}, {Kulkarni}, {Kumar},
  {Lee}, {Lenz}, {Littlefair}, {Ma}, {Macleod}, {Mastropietro}, {McCully},
  {Montagnac}, {Morris}, {Mueller}, {Mumford}, {Muna}, {Murphy}, {Nelson},
  {Nguyen}, {Ninan}, {N{\"o}the}, {Ogaz}, {Oh}, {Parejko}, {Parley}, {Pascual},
  {Patil}, {Patil}, {Plunkett}, {Prochaska}, {Rastogi}, {Reddy Janga},
  {Sabater}, {Sakurikar}, {Seifert}, {Sherbert}, {Sherwood-Taylor}, {Shih},
  {Sick}, {Silbiger}, {Singanamalla}, {Singer}, {Sladen}, {Sooley},
  {Sornarajah}, {Streicher}, {Teuben}, {Thomas}, {Tremblay}, {Turner},
  {Terr{\'o}n}, {van Kerkwijk}, {de la Vega}, {Watkins}, {Weaver}, {Whitmore},
  {Woillez}, {Zabalza}, \& {Astropy Contributors}}]{2018AJ....156..123A}
{Astropy Collaboration}, {Price-Whelan}, A.~M., {Sip{\H{o}}cz}, B.~M., {et~al.}
  2018, \aj, 156, 123, \dodoi{10.3847/1538-3881/aabc4f}

\bibitem[{{Bl{\"o}cker}(2001)}]{2001Ap&SS.275....1B}
{Bl{\"o}cker}, T. 2001, \apss, 275, 1.
\newblock \doarXiv{astro-ph/0102135}

\bibitem[{{Bowman} \& {Holdsworth}(2019)}]{2019A&A...629A..21B}
{Bowman}, D.~M., \& {Holdsworth}, D.~L. 2019, \aap, 629, A21,
  \dodoi{10.1051/0004-6361/201935640}

\bibitem[{{Coppejans} {et~al.}(2013){Coppejans}, {Gulbis}, {Kotze},
  {Coppejans}, {Worters}, {Woudt}, {Whittal}, {Cloete}, \&
  {Fourie}}]{2013PASP..125..976C}
{Coppejans}, R., {Gulbis}, A.~A.~S., {Kotze}, M.~M., {et~al.} 2013, \pasp, 125,
  976, \dodoi{10.1086/672156}

\bibitem[{{C{\'o}rsico} {et~al.}(2006){C{\'o}rsico}, {Althaus}, \& {Miller
  Bertolami}}]{2006A&A...458..259C}
{C{\'o}rsico}, A.~H., {Althaus}, L.~G., \& {Miller Bertolami}, M.~M. 2006,
  \aap, 458, 259, \dodoi{10.1051/0004-6361:20065423}

\bibitem[{{C{\'o}rsico} {et~al.}(2009){C{\'o}rsico}, {Althaus}, {Miller
  Bertolami}, {Gonz{\'a}lez P{\'e}rez}, \& {Kepler}}]{2009ApJ...701.1008C}
{C{\'o}rsico}, A.~H., {Althaus}, L.~G., {Miller Bertolami}, M.~M.,
  {Gonz{\'a}lez P{\'e}rez}, J.~M., \& {Kepler}, S.~O. 2009, \apj, 701, 1008,
  \dodoi{10.1088/0004-637X/701/2/1008}

\bibitem[{{C{\'o}rsico} {et~al.}(2019){C{\'o}rsico}, {Althaus}, {Miller
  Bertolami}, \& {Kepler}}]{2019A&ARv..27....7C}
{C{\'o}rsico}, A.~H., {Althaus}, L.~G., {Miller Bertolami}, M.~M., \& {Kepler},
  S.~O. 2019, \aapr, 27, 7, \dodoi{10.1007/s00159-019-0118-4}

\bibitem[{Craig {et~al.}(2017)Craig, Crawford, Seifert, Robitaille, Sipocz,
  Walawender, Vinícius, Ninan, Droettboom, Youn, Tollerud, Bray, walkerna22,
  Janga, stottsco, Günther, Rol, Bach, Bradley, Deil, Price-Whelan, Barbary,
  Horton, Schoenell, Nathan, Gasdia, Nelson, \&
  Streicher}]{matt_craig_2017_1069648}
Craig, M., Crawford, S., Seifert, M., {et~al.} 2017, astropy/ccdproc:
  v1.3.0.post1, v1.3.0.post1,  Zenodo, \dodoi{10.5281/zenodo.1069648}

\bibitem[{{Crowther} {et~al.}(1998){Crowther}, {De Marco}, \&
  {Barlow}}]{1998MNRAS.296..367C}
{Crowther}, P.~A., {De Marco}, O., \& {Barlow}, M.~J. 1998, \mnras, 296, 367,
  \dodoi{10.1046/j.1365-8711.1998.01360.x}

\bibitem[{{Dreizler} \& {Heber}(1998)}]{1998A&A...334..618D}
{Dreizler}, S., \& {Heber}, U. 1998, \aap, 334, 618

\bibitem[{{Gaia Collaboration} {et~al.}(2018){Gaia Collaboration}, {Brown},
  {Vallenari}, {Prusti}, {de Bruijne}, {Babusiaux}, {Bailer-Jones}, {Biermann},
  {Evans}, {Eyer}, {Jansen}, {Jordi}, {Klioner}, {Lammers}, {Lindegren},
  {Luri}, {Mignard}, {Panem}, {Pourbaix}, {Randich}, {Sartoretti}, {Siddiqui},
  {Soubiran}, {van Leeuwen}, {Walton}, {Arenou}, {Bastian}, {Cropper},
  {Drimmel}, {Katz}, {Lattanzi}, {Bakker}, {Cacciari}, {Casta{\~n}eda},
  {Chaoul}, {Cheek}, {De Angeli}, {Fabricius}, {Guerra}, {Holl}, {Masana},
  {Messineo}, {Mowlavi}, {Nienartowicz}, {Panuzzo}, {Portell}, {Riello},
  {Seabroke}, {Tanga}, {Th{\'e}venin}, {Gracia-Abril}, {Comoretto},
  {Garcia-Reinaldos}, {Teyssier}, {Altmann}, {Andrae}, {Audard},
  {Bellas-Velidis}, {Benson}, {Berthier}, {Blomme}, {Burgess}, {Busso},
  {Carry}, {Cellino}, {Clementini}, {Clotet}, {Creevey}, {Davidson}, {De
  Ridder}, {Delchambre}, {Dell'Oro}, {Ducourant},
  {Fern{\'a}ndez-Hern{\'a}ndez}, {Fouesneau}, {Fr{\'e}mat}, {Galluccio},
  {Garc{\'\i}a-Torres}, {Gonz{\'a}lez-N{\'u}{\~n}ez}, {Gonz{\'a}lez-Vidal},
  {Gosset}, {Guy}, {Halbwachs}, {Hambly}, {Harrison}, {Hern{\'a}ndez},
  {Hestroffer}, {Hodgkin}, {Hutton}, {Jasniewicz}, {Jean-Antoine-Piccolo},
  {Jordan}, {Korn}, {Krone-Martins}, {Lanzafame}, {Lebzelter}, {L{\"o}ffler},
  {Manteiga}, {Marrese}, {Mart{\'\i}n-Fleitas}, {Moitinho}, {Mora}, {Muinonen},
  {Osinde}, {Pancino}, {Pauwels}, {Petit}, {Recio-Blanco}, {Richards},
  {Rimoldini}, {Robin}, {Sarro}, {Siopis}, {Smith}, {Sozzetti}, {S{\"u}veges},
  {Torra}, {van Reeven}, {Abbas}, {Abreu Aramburu}, {Accart}, {Aerts},
  {Altavilla}, {{\'A}lvarez}, {Alvarez}, {Alves}, {Anderson}, {Andrei},
  {Anglada Varela}, {Antiche}, {Antoja}, {Arcay}, {Astraatmadja}, {Bach},
  {Baker}, {Balaguer-N{\'u}{\~n}ez}, {Balm}, {Barache}, {Barata}, {Barbato},
  {Barblan}, {Barklem}, {Barrado}, {Barros}, {Barstow}, {Bartholom{\'e}
  Mu{\~n}oz}, {Bassilana}, {Becciani}, {Bellazzini}, {Berihuete}, {Bertone},
  {Bianchi}, {Bienaym{\'e}}, {Blanco-Cuaresma}, {Boch}, {Boeche}, {Bombrun},
  {Borrachero}, {Bossini}, {Bouquillon}, {Bourda}, {Bragaglia}, {Bramante},
  {Breddels}, {Bressan}, {Brouillet}, {Br{\"u}semeister}, {Brugaletta},
  {Bucciarelli}, {Burlacu}, {Busonero}, {Butkevich}, {Buzzi}, {Caffau},
  {Cancelliere}, {Cannizzaro}, {Cantat-Gaudin}, {Carballo}, {Carlucci},
  {Carrasco}, {Casamiquela}, {Castellani}, {Castro-Ginard}, {Charlot},
  {Chemin}, {Chiavassa}, {Cocozza}, {Costigan}, {Cowell}, {Crifo}, {Crosta},
  {Crowley}, {Cuypers}, {Dafonte}, {Damerdji}, {Dapergolas}, {David}, {David},
  {de Laverny}, {De Luise}, {De March}, {de Martino}, {de Souza}, {de Torres},
  {Debosscher}, {del Pozo}, {Delbo}, {Delgado}, {Delgado}, {Di Matteo},
  {Diakite}, {Diener}, {Distefano}, {Dolding}, {Drazinos}, {Dur{\'a}n},
  {Edvardsson}, {Enke}, {Eriksson}, {Esquej}, {Eynard Bontemps}, {Fabre},
  {Fabrizio}, {Faigler}, {Falc{\~a}o}, {Farr{\`a}s Casas}, {Federici},
  {Fedorets}, {Fernique}, {Figueras}, {Filippi}, {Findeisen}, {Fonti},
  {Fraile}, {Fraser}, {Fr{\'e}zouls}, {Gai}, {Galleti}, {Garabato},
  {Garc{\'\i}a-Sedano}, {Garofalo}, {Garralda}, {Gavel}, {Gavras}, {Gerssen},
  {Geyer}, {Giacobbe}, {Gilmore}, {Girona}, {Giuffrida}, {Glass}, {Gomes},
  {Granvik}, {Gueguen}, {Guerrier}, {Guiraud}, {Guti{\'e}rrez-S{\'a}nchez},
  {Haigron}, {Hatzidimitriou}, {Hauser}, {Haywood}, {Heiter}, {Helmi}, {Heu},
  {Hilger}, {Hobbs}, {Hofmann}, {Holland}, {Huckle}, {Hypki}, {Icardi},
  {Jan{\ss}en}, {Jevardat de Fombelle}, {Jonker}, {Juh{\'a}sz}, {Julbe},
  {Karampelas}, {Kewley}, {Klar}, {Kochoska}, {Kohley}, {Kolenberg},
  {Kontizas}, {Kontizas}, {Koposov}, {Kordopatis}, {Kostrzewa-Rutkowska},
  {Koubsky}, {Lambert}, {Lanza}, {Lasne}, {Lavigne}, {Le Fustec}, {Le
  Poncin-Lafitte}, {Lebreton}, {Leccia}, {Leclerc}, {Lecoeur-Taibi},
  {Lenhardt}, {Leroux}, {Liao}, {Licata}, {Lindstr{\o}m}, {Lister}, {Livanou},
  {Lobel}, {L{\'o}pez}, {Managau}, {Mann}, {Mantelet}, {Marchal}, {Marchant},
  {Marconi}, {Marinoni}, {Marschalk{\'o}}, {Marshall}, {Martino}, {Marton},
  {Mary}, {Massari}, {Matijevi{\v{c}}}, {Mazeh}, {McMillan}, {Messina},
  {Michalik}, {Millar}, {Molina}, {Molinaro}, {Moln{\'a}r}, {Montegriffo},
  {Mor}, {Morbidelli}, {Morel}, {Morris}, {Mulone}, {Muraveva}, {Musella},
  {Nelemans}, {Nicastro}, {Noval}, {O'Mullane}, {Ord{\'e}novic},
  {Ord{\'o}{\~n}ez-Blanco}, {Osborne}, {Pagani}, {Pagano}, {Pailler},
  {Palacin}, {Palaversa}, {Panahi}, {Pawlak}, {Piersimoni}, {Pineau}, {Plachy},
  {Plum}, {Poggio}, {Poujoulet}, {Pr{\v{s}}a}, {Pulone}, {Racero}, {Ragaini},
  {Rambaux}, {Ramos-Lerate}, {Regibo}, {Reyl{\'e}}, {Riclet}, {Ripepi}, {Riva},
  {Rivard}, {Rixon}, {Roegiers}, {Roelens}, {Romero-G{\'o}mez}, {Rowell},
  {Royer}, {Ruiz-Dern}, {Sadowski}, {Sagrist{\`a} Sell{\'e}s}, {Sahlmann},
  {Salgado}, {Salguero}, {Sanna}, {Santana-Ros}, {Sarasso}, {Savietto},
  {Schultheis}, {Sciacca}, {Segol}, {Segovia}, {S{\'e}gransan}, {Shih},
  {Siltala}, {Silva}, {Smart}, {Smith}, {Solano}, {Solitro}, {Sordo}, {Soria
  Nieto}, {Souchay}, {Spagna}, {Spoto}, {Stampa}, {Steele},
  {Steidelm{\"u}ller}, {Stephenson}, {Stoev}, {Suess}, {Surdej}, {Szabados},
  {Szegedi-Elek}, {Tapiador}, {Taris}, {Tauran}, {Taylor}, {Teixeira},
  {Terrett}, {Teyssandier}, {Thuillot}, {Titarenko}, {Torra Clotet}, {Turon},
  {Ulla}, {Utrilla}, {Uzzi}, {Vaillant}, {Valentini}, {Valette}, {van Elteren},
  {Van Hemelryck}, {van Leeuwen}, {Vaschetto}, {Vecchiato}, {Veljanoski},
  {Viala}, {Vicente}, {Vogt}, {von Essen}, {Voss}, {Votruba}, {Voutsinas},
  {Walmsley}, {Weiler}, {Wertz}, {Wevers}, {Wyrzykowski}, {Yoldas},
  {{\v{Z}}erjal}, {Ziaeepour}, {Zorec}, {Zschocke}, {Zucker}, {Zurbach}, \&
  {Zwitter}}]{2018A&A...616A...1G}
{Gaia Collaboration}, {Brown}, A.~G.~A., {Vallenari}, A., {et~al.} 2018, \aap,
  616, A1, \dodoi{10.1051/0004-6361/201833051}

\bibitem[{{Grauer} {et~al.}(1987{\natexlab{a}}){Grauer}, {Bond}, {Green}, \&
  {Liebert}}]{1987fbs..conf..231G}
{Grauer}, A.~D., {Bond}, H.~E., {Green}, R.~F., \& {Liebert}, J.
  1987{\natexlab{a}}, in IAU Colloq. 95: Second Conference on Faint Blue Stars,
  ed. A.~G.~D. {Philip}, D.~S. {Hayes}, \& J.~W. {Liebert}, 231--236

\bibitem[{{Grauer} {et~al.}(1987{\natexlab{b}}){Grauer}, {Bond}, {Liebert},
  {Fleming}, \& {Green}}]{1987ApJ...323..271G}
{Grauer}, A.~D., {Bond}, H.~E., {Liebert}, J., {Fleming}, T.~A., \& {Green},
  R.~F. 1987{\natexlab{b}}, \apj, 323, 271, \dodoi{10.1086/165824}

\bibitem[{{Green} {et~al.}(1986){Green}, {Schmidt}, \&
  {Liebert}}]{1986ApJS...61..305G}
{Green}, R.~F., {Schmidt}, M., \& {Liebert}, J. 1986, \apjs, 61, 305,
  \dodoi{10.1086/191115}

\bibitem[{{Herwig}(2001)}]{2001Ap&SS.275...15H}
{Herwig}, F. 2001, \apss, 275, 15.
\newblock \doarXiv{astro-ph/9912353}

\bibitem[{Hunter(2007)}]{Hunter:2007}
Hunter, J.~D. 2007, Computing in Science \& Engineering, 9, 90,
  \dodoi{10.1109/MCSE.2007.55}

\bibitem[{{Iben} {et~al.}(1983){Iben}, {Kaler}, {Truran}, \&
  {Renzini}}]{1983ApJ...264..605I}
{Iben}, I., J., {Kaler}, J.~B., {Truran}, J.~W., \& {Renzini}, A. 1983, \apj,
  264, 605, \dodoi{10.1086/160631}

\bibitem[{{Lenz} \& {Breger}(2005)}]{2005CoAst.146...53L}
{Lenz}, P., \& {Breger}, M. 2005, Communications in Asteroseismology, 146, 53,
  \dodoi{10.1553/cia146s53}

\bibitem[{{Loumos} \& {Deeming}(1978)}]{1978Ap&SS..56..285L}
{Loumos}, G.~L., \& {Deeming}, T.~J. 1978, \apss, 56, 285,
  \dodoi{10.1007/BF01879560}

\bibitem[{{Miller Bertolami} \& {Althaus}(2006)}]{2006A&A...454..845M}
{Miller Bertolami}, M.~M., \& {Althaus}, L.~G. 2006, \aap, 454, 845,
  \dodoi{10.1051/0004-6361:20054723}

\bibitem[{{Quirion} {et~al.}(2004){Quirion}, {Fontaine}, \&
  {Brassard}}]{2004ApJ...610..436Q}
{Quirion}, P.~O., {Fontaine}, G., \& {Brassard}, P. 2004, \apj, 610, 436,
  \dodoi{10.1086/421447}

\bibitem[{{Quirion} {et~al.}(2007){Quirion}, {Fontaine}, \&
  {Brassard}}]{2007ApJS..171..219Q}
---. 2007, \apjs, 171, 219, \dodoi{10.1086/513870}

\bibitem[{{Schuh} {et~al.}(2000){Schuh}, {Dreizler}, {Deetjen}, {Heber}, \&
  {Geckeler}}]{2000BaltA...9..395S}
{Schuh}, S., {Dreizler}, S., {Deetjen}, J.~L., {Heber}, U., \& {Geckeler},
  R.~D. 2000, Baltic Astronomy, 9, 395, \dodoi{10.1515/astro-2000-0306}

\bibitem[{{Sowicka} {et~al.}(2018){Sowicka}, {Handler}, \&
  {Jones}}]{2018MNRAS.479.2476S}
{Sowicka}, P., {Handler}, G., \& {Jones}, D. 2018, \mnras, 479, 2476,
  \dodoi{10.1093/mnras/sty1660}

\bibitem[{{Starrfield} {et~al.}(1984){Starrfield}, {Cox}, {Kidman}, \&
  {Pesnell}}]{1984ApJ...281..800S}
{Starrfield}, S., {Cox}, A.~N., {Kidman}, R.~B., \& {Pesnell}, W.~D. 1984,
  \apj, 281, 800, \dodoi{10.1086/162158}

\bibitem[{{Starrfield} {et~al.}(1983){Starrfield}, {Cox}, {Hodson}, \&
  {Pesnell}}]{1983ApJ...268L..27S}
{Starrfield}, S.~G., {Cox}, A.~N., {Hodson}, S.~W., \& {Pesnell}, W.~D. 1983,
  \apjl, 268, L27, \dodoi{10.1086/184023}

\bibitem[{{Steininger} {et~al.}(2003){Steininger}, {Paparo}, {Viraaghalmy},
  {Zsuffa}, \& {Breger}}]{2003ASPC..292..237S}
{Steininger}, B., {Paparo}, M., {Viraaghalmy}, G., {Zsuffa}, D., \& {Breger},
  M. 2003, in Astronomical Society of the Pacific Conference Series, Vol. 292,
  Interplay of Periodic, Cyclic and Stochastic Variability in Selected Areas of
  the H-R Diagram, ed. C.~{Sterken}, 237

\bibitem[{{Werner}(2001)}]{2001Ap&SS.275...27W}
{Werner}, K. 2001, \apss, 275, 27

\bibitem[{{Werner} \& {Heber}(1991)}]{1991A&A...247..476W}
{Werner}, K., \& {Heber}, U. 1991, \aap, 247, 476

\bibitem[{{Werner} {et~al.}(2008){Werner}, {Rauch}, {Reiff}, \&
  {Kruk}}]{2008ASPC..391..109W}
{Werner}, K., {Rauch}, T., {Reiff}, E., \& {Kruk}, J.~W. 2008, in Astronomical
  Society of the Pacific Conference Series, Vol. 391, Hydrogen-Deficient Stars,
  ed. A.~{Werner} \& T.~{Rauch}, 109.
\newblock \doarXiv{0710.4506}

\bibitem[{{Winget} \& {Kepler}(2008)}]{2008ARA&A..46..157W}
{Winget}, D.~E., \& {Kepler}, S.~O. 2008, \araa, 46, 157,
  \dodoi{10.1146/annurev.astro.46.060407.145250}

\end{thebibliography}
\bibliographystyle{aasjournal}



\end{document}